\documentclass{aa}
\usepackage{graphicx}
\usepackage{txfonts}
\begin{document}

   \title{Eclipsing binary statistics -\\
   theory and observation}

   \subtitle{}

   \author{Staffan S\"oderhjelm
              \and
          Johann Dischler
          }

        \institute{Lund Observatory,
              Box 43, SE-22100 Lund, Sweden \\
              \email{staffan@astro.lu.se, johann@astro.lu.se}}

   \offprints{Johann Dischler}

   \date{Received 15 December 2004 / Accepted 20 May 2005}

   \abstract{The expected distributions of eclipse-depth versus
   period for eclipsing binaries of different luminosities are derived
from large-scale population synthesis experiments. Using the rapid
Hurley et al. BSE binary evolution code, we have evolved several
hundred million binaries, starting from various simple input
distributions of masses and orbit-sizes. Eclipse probabilities and
predicted distributions over period and eclipse-depth (P/$\Delta
m$) are given in a number of main-sequence intervals, from O-stars
to brown dwarfs. The comparison between theory and Hipparcos
observations shows that a standard (Duquennoy\& Mayor) input
distribution of orbit-sizes (a) gives reasonable numbers and
P/$\Delta m$-distributions, as long as the mass-ratio distribution
is also close to the observed flat ones. A random pairing model,
where the primary and secondary are drawn independently from the
same IMF, gives more than an order of magnitude too few eclipsing
binaries on the upper main sequence. For a set of eclipsing
OB-systems in the LMC, the observed period-distribution is
different from the theoretical one, and the input orbit
distributions and/or the evolutionary environment in LMC has to be
different compared with the Galaxy. A natural application of these
methods are estimates of the numbers and properties of eclipsing
binaries observed by large-scale surveys like Gaia.

   }

   \titlerunning{Eclipsing binary statistics - Theory and observation}
   \authorrunning{S. S\"oderhjelm \& J. Dischler}
   \maketitle
%
\section{Introduction}

Eclipsing binaries are important tools for investigating stellar
parameters, but with periods and eclipse-depths taken as provided
by Nature. Some pioneering statistics for the period-distributions
for eclipsing binaries of different types were given by Farinella
\& Paolicchi (\cite{Farinella1}) and Antonello et al.
(\cite{Farinella2}), but the distribution of the eclipse-depths
has less often been studied in its own right. An important reason
is that such statistics are very incomplete. Eclipsing binaries
have been discovered from non-systematic searches, and their
physical natures are too diverse for meaningful statistics.
Forthcoming large-scale surveys like Gaia will however observe
large numbers of eclipsing binaries, and it is clearly interesting
to be able to give at least crude estimates of the numbers
expected in different parts of the HR diagram.

The two most important observable parameters for an eclipsing
binary are the period and the (maximum) eclipse depth. We also
generally know at least an approximate absolute magnitude or mass
for the system. For all-sky surveys like Hipparcos or Gaia, we
have an 'all-age/all-mass' sample, and as a first theoretical
study, it seems reasonable to evolve time- and space-invariable
initial distributions of masses and orbits during a 12 Gyr
evolution of a Galactic disk. We will thus use a large sample of
binaries, trying to reproduce their present characteristics by a
large-scale population synthesis. The eclipse probability depends
mainly on the ratio between the stellar radii and the orbit size,
and we get appreciable numbers of eclipsing binaries only at
periods of weeks or days. Because of orbital evolution, however,
such a short-period system may have started out in a much wider
orbit, and to keep this possibility, the population synthesis is
started with a very wide distribution of orbit-sizes (a), and
typically millions of original binaries are needed to get a few
hundred eclipsing pairs.

For all the interesting (short) periods, the evolution includes
episodes of mass-transfer or a common-envelope phase, and the
parameters of the final binaries cannot be obtained without
following this evolution in detail. The rapid binary evolution
code by Hurley et al. (\cite{Hurley2}), hereafter BSE, is here a
fundamental tool. Using large numbers of analytic interpolation
formulae, the BSE code summarizes and extends data from a large
number of detailed evolution models in a unified structure where a
given binary with arbitrary mass, metallicity and orbit can be
followed in time. The rapid results obtainable with BSE enables
the large-scale studies needed to get sufficient numbers of
eclipsing pairs, but the realism of the final data also depends
crucially on the accuracy and parameter choices in the BSE code.

The basic assumptions and parameters used in our theoretical
investigations are described in Sects. 2 and 3. Because the
eclipse probabilities vary mostly with the stellar radii, it is
useful to combine data in smaller parts of the HR diagram, where
the radii are fairly similar. This binning along the main sequence
is described in Sect. 4, and in Sect. 5 we present some typical
results. Section 6 explores the sensitivity of these results to
various changes of the input assumptions, before the theory is
compared with observed data in Sects. 7 (Hipparcos) and 8 (LMC).
Section 9 gives our preliminary conclusions, and the Appendix
shows some sample tables.

\section{Population synthesis for binaries}
We try to reproduce galactic disk data starting from simple
original distributions of masses and orbits. We simply assume a
constant Star Formation Rate (0-12 Gyr), a time-independent IMF
(Kroupa \cite{Kroupa}), with
\begin{equation}
\xi(m) \varpropto m^{-\alpha_i},
\end{equation}
and with
\begin{eqnarray}
      \alpha_0 &=& 0.3, \qquad 0.03 < m/M_{\odot} < 0.08 \\
      \alpha_1 &=& 1.8, \qquad 0.08 < m/M_{\odot} < 0.50 \nonumber \\
          \alpha_2 &=& 2.7, \qquad 0.50 < m/M_{\odot} < 1.00 \nonumber \\
          \alpha_3 &=& 2.3, \qquad 1.00 < m/M_{\odot} < 50.0 \nonumber
\end{eqnarray}
The metallicity distribution in the synthesis program is
age-independent and uniform in [Fe/H], and several runs with
different upper and lower limits are combined in order to mimic
the observed distributions. For the galactic samples, 10\% of the
[Fe/H]-values are in the interval [0.1,0.23], 70\% in the interval
[-0.3,0.1] and 20\% in the interval [-0.8,-0.3]. This gives a
total distribution as broad as e.g. the one shown in Nordstr\"om
et al. (\cite{Nordstrom}).

To allow later scalings to different multiplicity fractions, we
either make all the IMF mass fragments into single stars, or all
to binaries. The binary runs are of most interest to us, and we
have to assume a distribution f(a,q,e) of orbit-sizes, initial
mass-ratios and initial eccentricities. Even if it is clear that
this is in general not separable, we make the simplistic first
assumption $f(a,q,e)=f(a)\times f(q)\times f(e)$.

The q-distribution (with $q<1$ for the ratio of the originally
smaller to the larger mass) is mostly the rather ad hoc compromise
function
\begin{equation} 
   f(q)= 0.446 \times [1+2  n(q-0.2,0.3)+2  n(q-1.0,0.05)]
\end{equation}
with $n(m,s)=exp(-m^2/(2s^2))$. The narrow peak at $q=1$ is
reduced as compared with S\"oderhjelm (\cite{Soderhjelm}), because
it is not seen by e.g. Mazeh et al. (\cite{Mazeh}), or as a much
broader feature by Halbwachs et al. (\cite{Halbwachs}). Its total
contribution to the above $f(q)$ is only about 6\%, and it is
relatively unimportant in the present study. For comparison, we
have used also the gaussian distribution given by Duquennoy \&
Mayor (\cite{Duquennoy}), as well as a random pairing scheme, each
component of the binary picked independently from the IMF and
paired randomly. See Sect. 6.

The distribution of the semi-major axes of the binaries is usually
taken as
\begin{equation}
f(\lg a) = 0.269 \times n(\lg a-1.5,1.5) \qquad -3 < \lg a [a.u.]
< 5
\end{equation}
which is close to the (period-)relation given by Duquennoy \&
Mayor (\cite {Duquennoy}). Some other choices are tested in Sect.
6.

The eccentricities are distributed thermally for the largest
orbits,
\begin{equation}
f(e) \varpropto 2e \qquad a > 1000\; \mathrm{a.u.}
\end{equation}
turning smoothly into a uniform  distribution at $a < 10\;
\mathrm{a.u.}$ The tidal evolution built into the BSE-code usually
produces a rapid circularization for periods below about 10 days.

The BSE routines include a number of (sometimes poorly known)
parameters, but to keep the amount of computations reasonable, we
used only the default values given in Table 3 in Hurley et al.
(\cite {Hurley2}). If these parameters can be trusted, the BSE
code gives for each input binary the actual masses, temperature
and luminosities of the components, as well as the orbit size and
shape, even when the evolution includes episodes of mass-transfer
and common-envelope evolution. A minor problem occurs at brown
dwarf masses, since the BSE models are accurate only down to a
bona fide stellar mass-limit around 0.08 $M_{\sun}$. For any
component of lower mass, we added an extra grid with data due to
Baraffe et al. (\cite{Baraffe}), giving more realistic cooling
brown dwarfs.

\begin{figure}
   \centering
   \includegraphics[width=8.5cm,angle=0]{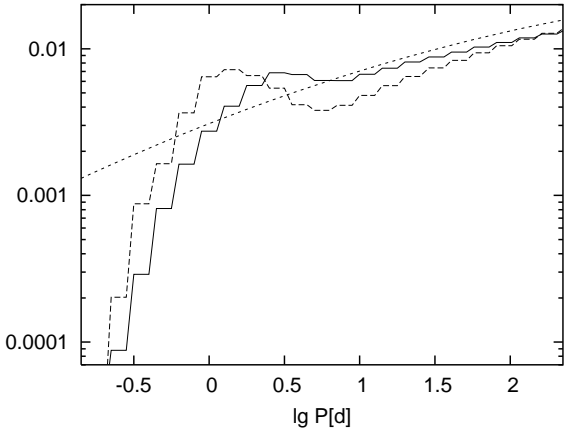}
      \caption{The short-period part of the period-distribution for systems with
      F- or G-star primaries. The full line shows the BSE calculation, the dashed one
      the S+S approximation, and the dotted line the smooth input distribution.
              }
         \label{figadistr}
   \end{figure}

For orbits with periods above about 5 years, each component in the
binary evolves independently, and we can make a binary from two
single-star models, using analytical approximations to the
single-star evolutionary tracks given by Hurley et al.
(\cite{Hurley1}), and again with the Baraffe brown dwarf
extension. As a check of the importance of the full BSE models, we
have used in parallel this S+S approximation. There is the
non-trivial problem of how to evolve the orbit (size and
eccentricity) when one or both of the components lose mass. We
used here a simplistic prescription with the semi-major axis
inversely proportional to the total remaining mass in the system,
and with a circularization time-scale given by $\lg t_{\rm
circ}[yr]=23+10\lg a[a.u.]-8\lg R[Rsun]$ (where R is the radius of
the largest star in the system). When any of the components
reached its Roche-lobe, the system was removed. Still, this S+S
approach gives a qualitatively similar (post-evolution)
a-distribution, with a hump before an obvious contact-limit. Fig
\ref{figadistr} shows the short-period end of the a-distributions
for the FG-bin (see Sect. 4), as calculated by BSE and by the S+S
approximation. In this and other cases, the S+S model gives
(counter-intuitively) more short-period pairs than the more
realistic BSE calculations.

Both for the BSE and S+S runs, there is the standard problem of
going from theoretical ($M_{\rm bol},T_{\rm eff}$) to observable
($M_V$, V-I) coordinates, and again only illustrative
transformations are used, based on the data in Flower
(\cite{Flower}) but with misprint corrections and extensions by P.
Nurmi (private comm.) and S. S\"oderhjelm. For cool stars, these
transformations are admittedly rather poor, but their influence on
the $\Delta m$-statistics is very indirect, and more elaborate
models are not called for in this first study.

   \begin{figure}
   \centering
   \includegraphics[width=9.2cm,angle=0]{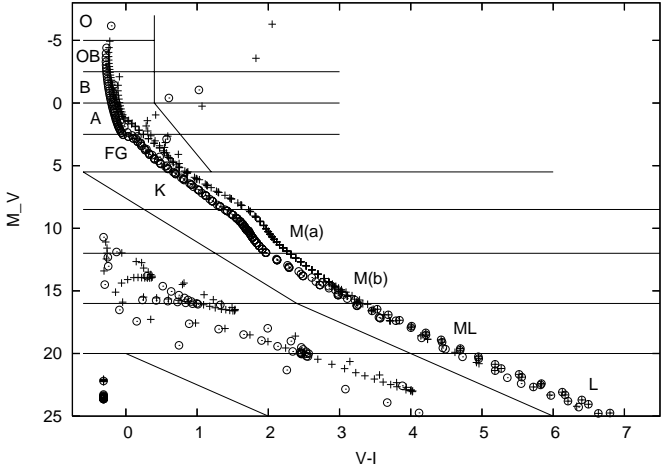}
      \caption{The binning in the CM-diagram, with the main-sequence bins
      labelled with approximate spectral types. About 100 stellar models (0.03-45 Msun) are plotted at age 10, 100, 1000, 3000 and 10000 Myr,
      for [Fe/H]=0 (crosses) or [Fe/H]=-2 (open circles), showing
      the main locations for the MS and WD sequences.
              }
         \label{fig_HRbins}
   \end{figure}

\section{Eclipse statistics}

The population synthesis gives a number of binaries with known
properties, among them the orbit size and period and the stellar
radii and luminosities. The radii relative to the orbit-size $r_s
= R_s/a$ and $r_g = R_g/a$ (where s and g stand for smaller and
greater) are the interesting parameters, together with the
relative (V-band) luminosities $L_s$ and $L_g$, normalized to
$L_s+L_g=1$. It is easy to show that the probability of an eclipse
of any depth is simply $P_{e_0} = r_s+r_g$, but we want to study
the probability of an eclipse of any specified (magnitude) depth.
In this first study, we have neglected all light-curve
complications due to tidal deformation and/or reflection effects,
and in the same vein, we also assume zero limb-darkening.

For this simple case, the light-curve can be calculated from the
radii, luminosities and orbital inclination, as shown e.g. by
Irwin (\cite{Irwin}). Although the maximum (primary) eclipse depth
$\Delta m$ is a well-defined function of the inclination $i$,
iteration is needed for the inverse problem of deriving the
function $i(\Delta m)$. With a random distribution of
inclinations, the probability $P_e(\Delta m)$ of a primary eclipse
deeper than $\Delta m$ is now simply $\cos(i(\Delta m))$. As
expected, the inclination at grazing eclipse ($\Delta m = \epsilon
\approx 0$) is then $i(\epsilon)=\arccos(r_s+r_g)$, giving
$P_e(\epsilon)=P_{e_0}$.

For main sequence stars of roughly constant size, the $P_e$-values
decrease as $1/a$, or as $P^{-2/3}$. In practice, however, a
longer period also means narrower (easily missed) eclipses, and
the observed bias towards smaller periods is more pronounced. To
account for this effect, a nonzero $P_e$-value is only calculated
if the $\Delta m$ is larger than 0.1 magnitude for at least 5 \%
of the orbit. As shown e.g. in S\"oderhjelm (\cite{sod05}), such
discovery criteria have a large impact on the derived results. In
the present study, we have used the 5\%/0.1 mag limits throughout,
except for a comparison in Sect. 7.

Mean values for these eclipse probabilities are meaningful only
for roughly similar systems, meaning roughly similar radii,
luminosities and orbit sizes. In practice, we have divided the HR
diagram in a number of bins of absolute magnitude and color, and
for the stars in each bin, we then further subdivide according to
orbital period and $\Delta m$. The final results are histograms
giving the mean $P_e(\Delta m,P)$ in each separate HR-diagram bin.
These eclipse probabilities are interesting in themselves, and one
may compare e.g. the BSE and S+S results.

However, for comparisons with observed eclipsing binaries, another
type of probability is more interesting, namely the fraction of
eclipsing binaries relative to all stars (binary and single) in
the same HR diagram bin. In order to derive such a normalized
probability $O_e(\Delta m,P)$, we multiply the corresponding
$P_e(\Delta m,P)$ by the number of binaries in that particular
period range and divide by the total number of binaries (of all
periods) and single stars in the specified HR diagram bin. The
$O_e$ values are sensitive to the period-distribution of the
binaries, and also to the absolute frequency of double relative to
single stars. In our model, we assume a nominal 80 \% duplicity
fraction, in the sense that only 20 \% of the mass fragments from
the IMF are kept as single stars, while 80\% are split into
doubles. Because a single has higher mass than the primary in a
double created from the same mass fragment, the duplicity fraction
when comparing singles and primaries of the same mass is lower
(and variable along the main sequence).
\begin{table}
  \centering

\begin{tabular}{|l|cr|}

  \hline

  Bin-id & $M_V$-interval & $<M_V>$  \\
      \hline
  O & $<$ -5       &  -5.31   \\
  OB & -5.0 - -2.5 & -3.04   \\
  B & -2.5 - 0.0   & -0.55   \\
  A &  0.0 - 2.5   &  1.79   \\
  FG & 2.5 - 5.5   & 4.35  \\
  K & 5.5 - 8.5    & 7.15   \\
  M(a) & 8.5 - 12.0 &  10.70  \\
  M(b) & 12.0 - 16.0 & 13.99   \\
  ML & 16.0 - 20.0 & 17.71 \\
  L &  20.0 - 25.0 & 22.44 \\
  \hline
\end{tabular}
  \caption{The magnitude limits for the main-sequence bins, and the median absolute magnitude
  for the stars in the bin, for the standard input assumptions.}\label{bintab}
\end{table}

\section{Covering the MS}

The overall $M_V/(V-I)$ color-magnitude diagram for all
metallicities and all ages shows a very wide main sequence, and
the binning selected is similarly coarse. Fig. \ref{fig_HRbins}
shows the separation into 22 regions, together with some typical
isochrones. The color-boundaries are designed for a basic
separation into MS, giant and wd parts, and on the MS, the
$M_V$-intervals can be (illustratively) translated to spectral
type. The naming-convention is shown in Fig. \ref{fig_HRbins} and
Table \ref{bintab}.

The simplistic approach of making a single run with the full IMF
and the full 0-12 Gyr age range would require an inconvenient
number of simulated stars, and would still give in the end very
few upper MS eclipsing binaries. Instead, we have made a series of
partial runs, each covering mass-sums from a lower limit to the
BSE maximum (around 45 $M_{\odot}$), and with a maximum age that
allows the lowest-mass pairs to evolve past the MS stage. In this
way, most of the objects at any time populating this MS-bin can be
generated (allowing for massive binaries evolving downwards, but
with no expectation that lower-mass pairs may ever become brighter
than a single star with their combined mass). For the upper MS,
one or two of these partial runs were designed to populate each
bin, and in order to have large samples of close binaries, the
total numbers (of wider systems) were very large. As described in
Sect. 2, each 'run' was also routinely split in three or more
partial ones with different ranges of metallicity)

For example, the O bin was created from a total of 19 million
systems with mass-sums larger than 16 $M_{\odot}$ and ages 0-12
Myr. For the A bin, 23 million simulated systems with mass-sums
above 1.35 $M_{\odot}$ were used, with ages 0-3000 Myr. In total,
we spent about 2500 (2 Ghz Pentium 4) CPU-hours to simulate about
250 million binaries. The magnitude limits defining the different
bins are given in Table \ref{bintab}. Because of the IMF slope,
the median absolute magnitude in a bin is not central between
these bin-limits, but shifted towards the faint edge, or, for the
least massive dwarfs, towards the bright edge instead.

   \begin{figure}
   \centering
   \mbox{
   \includegraphics[width=9cm]{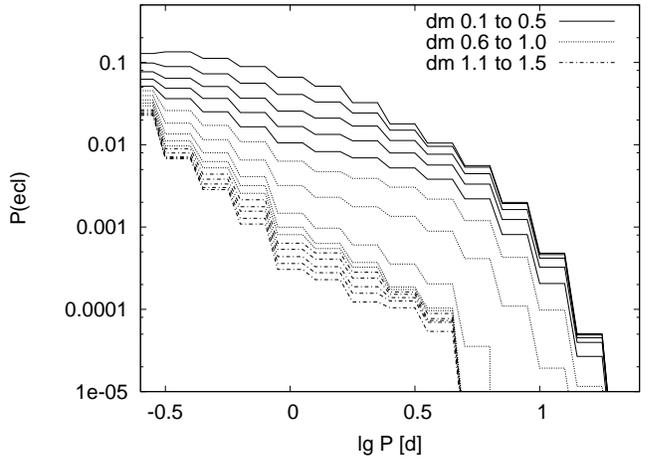}
   }
   \caption{The eclipse probabilities $P_e(\Delta m,P)$ for the FG
            bin, as a function of the (log) period in days. Each of the lines
            (solid from 0.1 to 0.5 mag, dashed from 0.6 to 1.0,
            and dot-dashed from 1.1 to 1.5 mag) corresponds
            to a certain maximum eclipse-depth $\Delta m$.}
   \label{fig_FGstars_Pe}
   \end{figure}

   \begin{figure}
   \centering
   \mbox{
   \includegraphics[width=9cm]{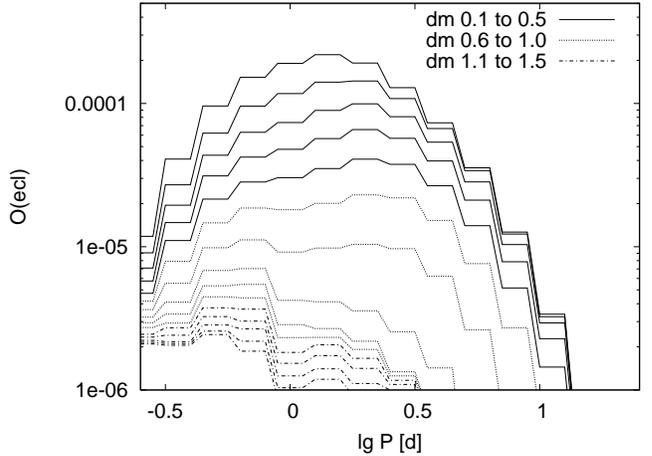}
   }
   \caption{The $O_e(\Delta m,P)$-curves, giving the fraction of eclipsing systems
            (per 0.3 dex in lg P) relative to all stars, again for the
            FG bin. Same key for the $\Delta m$:s as in
             Fig. \ref{fig_FGstars_Pe}.}
   \label{fig_FGstars_O}
   \end{figure}
   \begin{figure}
   \centering
   \mbox{
   \includegraphics[width=9cm]{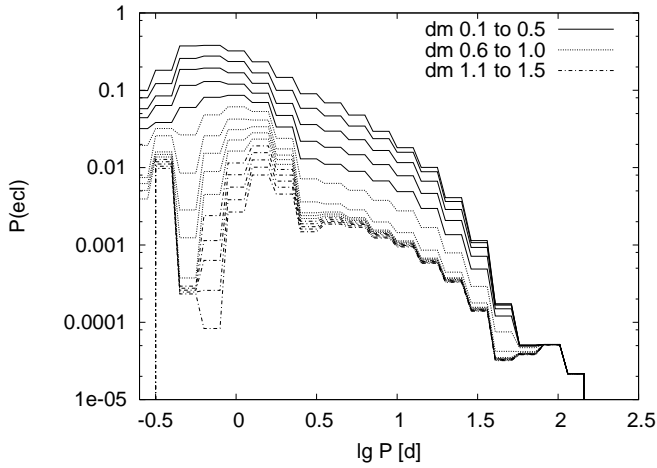}
   }
   \caption{The eclipse probabilities $P_e(\Delta m,P)$ for B systems. Because more systems have
   undergone mass transfer, the details are more complex than in Fig. \ref{fig_FGstars_Pe}.}
   \label{fig_Bstars_Pe}
   \end{figure}

\section{Typical results}
The basic output from our studies are histograms (with a 0.15 dex
binning in $\lg P$) giving the mean probability of observing an
eclipse of \emph{at least} a (maximum) depth $\Delta m$. Fig
\ref{fig_FGstars_Pe} shows a plot of these data for the FG-bin,
with the 0.1 mag curve at the top and the 1.5 mag curve at the
bottom. Because the probabilities for deep eclipses are so small,
a logarithmic scale is necessary to show the details. For the
$O_e$-plot in Fig \ref{fig_FGstars_O}, we have taken the summed
$P_e$-values in each period bin, and divided by the total number
of (singles and binaries of all periods) in the whole FG-bin.
(More specifically, the $O_e$-values are defined per 0.3 dex
period-interval and not the 0.15 dex histogram spacing. We have
also used the 80\% duplicity defined in Sect. 3). The $O_e$-curves
decrease at short periods because the number of short-period pairs
decreases (cf. Fig \ref{figadistr}), although each such system has
a high probability of eclipses. The decrease towards larger
periods is shaped also by the assumed discovery-criterion, see
Sect. 7.

The deeper eclipse for any eclipsing binary is due to the cooler
star eclipsing the hotter. For unevolved main-sequence stars, a
hotter star is always larger, and the partial primary eclipses can
be shown to be never deeper than the 0.75 magnitude limiting case
when two equal stars mutually cover each other. A deeper primary
eclipse needs a more complex evolution with Roche-lobe overflow
and mass-transfer producing Algol systems with subgiant
secondaries that may totally eclipse their hotter primaries.

Although typically less than 1\% of the eclipsing FG systems show
1.5 mag eclipses, the ratio at the shortest periods is closer to
20\%, showing a much higher proportion of such Algol systems. The
situation is even more complex for massive systems, as seen e.g.
in Fig \ref{fig_Bstars_Pe} for the B bin. There are now
characteristic humps with a high probability of deep eclipses at
about 2 day and 0.4 day orbital periods. The longer-period hump is
visible from A- to OB-stars, although it shifts successively
towards larger periods for the larger masses. It would be
interesting to investigate further which kinds of progenitor
systems produce these deeply eclipsing systems, and what
parameters influence their periods, but we have not yet tried to
do so. As will be seen in Sect. 8, the hump period seems to be
larger in the LMC than in the Galaxy, and more experimenting will
be needed to understand all the factors involved.

\begin{figure}
   \centering
   \includegraphics[width=9cm]{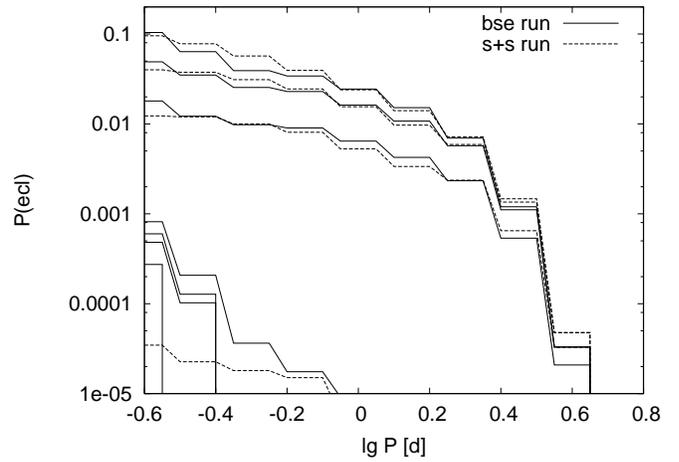}
   \caption{Comparison of $P_e$-data for low-mass M-stars (M(b) bin), as calculated
   with BSE (solid lines) and with a simplified single+single model (dashed
   lines). Values are given for (from above) $\Delta m$=0.2, 0.4, 0.6,...1.4, but
            the S+S curves for $\Delta m > 0.8$ disappear below
            the plot area.}
   \label{fig_Mstars_Pe}
\end{figure}

Our S+S binaries should be identical to the BSE ones as long as
there is no mass-transfer between the components. In practice,
there are differences already in the orbit evolution when a
substantial fraction of the mass is lost from the system in the
giant stage. When both stars are on the main sequence, however,
the results are quite similar. The low-mass M(b)-stars are
expected to remain on the MS for more than 12 Gyr, and as seen in
Fig \ref{fig_Mstars_Pe}, the $\Delta m$-curves for 0.2, 0.4 and
0.6 mag are more or less equal for S+S and BSE. In the S+S model,
the 0.8 mag curve is also almost at zero (by the '0.75-effect'),
but with BSE, some originally higher-mass binaries have after a
common-envelope phase actually become short-period systems with wd
primaries and M-type secondaries. Going up the main-sequence, the
differences between the S+S and BSE models increase, in the sense
that only BSE can create the deeply eclipsing Algol systems.

\begin{figure}
   \centering
   \includegraphics[width=9cm]{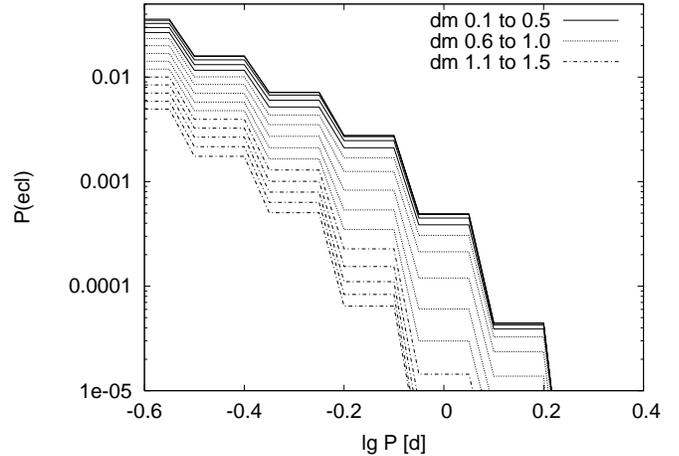}
   \caption{The $P_e(\Delta m,P)$-curves for brown dwarfs (ML and L bins).
   Standard (Fig. \ref{fig_FGstars_Pe}) labelling of the $\Delta m$ curves.
            }
   \label{fig_brdw}
\end{figure}

An interesting feature of bona-fide brown dwarfs is that their
radii remain rather constant while they cool, making it possible
to find totally eclipsing systems where the eclipsing star is much
fainter than the eclipsed one. The BSE data are in this case
identical to the S+S ones, in both cases based on our
implementation of the Baraffe et al. (\cite{Baraffe}) data. The
$P_e$-curves shown in Fig \ref{fig_brdw} could be extended at a
similar equidistant spacing down to quite large $\Delta m$ (but
more systems would have to be simulated to get good enough
statistics). The curves for 0.1 to 0.4 mag eclipse depth more or
less merge, because of the improbability of close to grazing
eclipses for very small stars.

\begin{figure}
   \centering
   \mbox{
   \includegraphics[width=9cm]{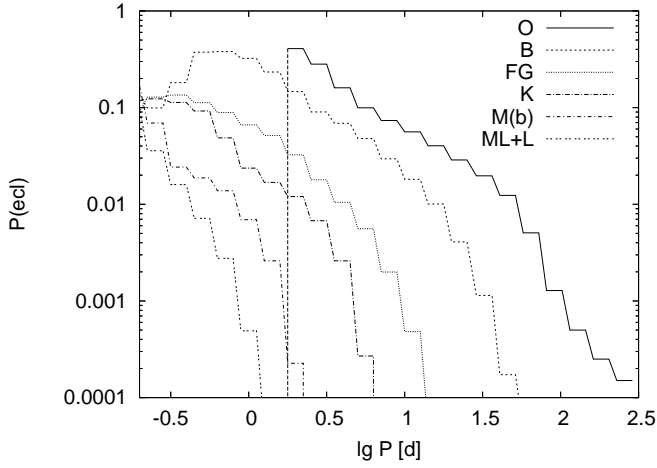}
   }
  \caption{The eclipse probabilities $P_e(0.1,P)$ for a number of different
  MS bins, as indicated in the legend.}
   \label{fig_AllTypes_dm01}
   \end{figure}

In Fig \ref{fig_AllTypes_dm01}, we show only the largest
eclipse-probabilities ($\Delta m>0.1$ mag), but for several
absolute magnitude bins at once. There is now a very definite
trend with higher probabilities (at longer periods) for brighter
stars, because they are larger. For the O systems, the
short-period contact-limit is very apparent, with no room for
shorter period systems with these large radii.

For a fuller sample of the data, see Appendix A.

\section{Varying the input parameters}
All the above results are for a single set of input parameters.
Even if the assumptions are reasonable, it is obviously very
important to know how a different choice would affect the results.
Ideally, one would like to vary one parameter at a time, redo all
calculations, and see the differential change in the output. With
tens of parameters to look at, the amount of computation is
prohibitive, and for simplicity, we have not tried to vary the
parameters in BSE. Foremost among the poorly-known ones is the
notorious 'common-envelope efficiency parameter', which is kept at
$\alpha_{\rm CE}=3.0$. We have concentrated instead on the input
distributions of orbit sizes and mass-ratios, and on the effects
of the metallicity.

\begin{figure}
   \centering
   \includegraphics[width=9cm]{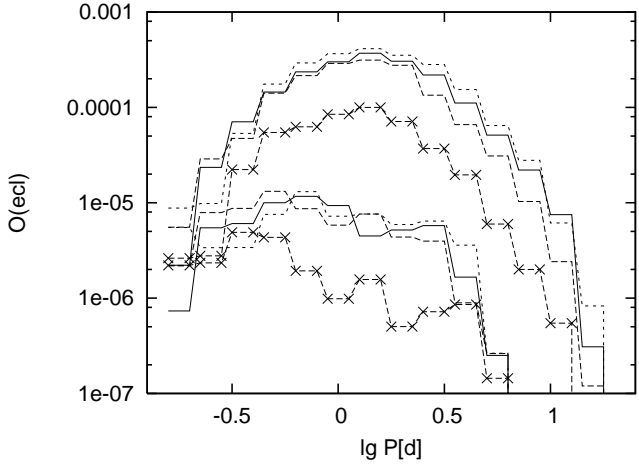}
   \caption{Different versions of the $O_e$-curves for $\Delta m>$0.1 and 0.8, for the FG bin.
            The three closely spaced curves give results using the standard q-distribution (full line), the
            (DM) Q1 version (dashed), or the (uniform) Q2 version (dotted line). A factor 5 below
            the other curves (dashed line plus crosses) are the ones calculated with the random pairing
            Q3 input.}

   \label{figvarq}
\end{figure}
\begin{figure}
   \centering
   \includegraphics[width=9cm]{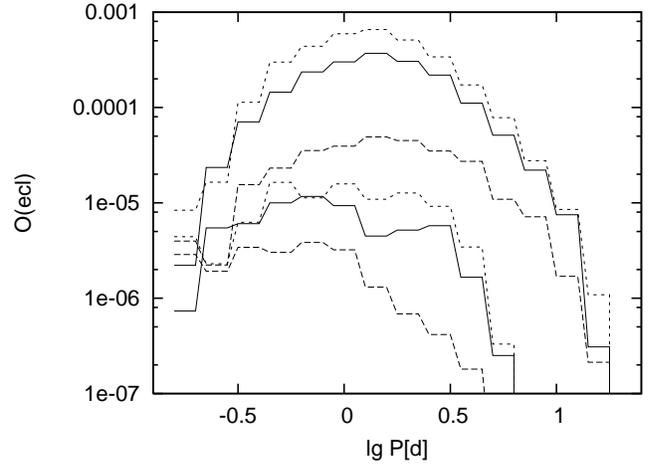}
   \caption{Different versions of the $O_e$-curves for $\Delta m>$0.1 and 0.8, for the FG bin.
            The full lines are for the standard a-distribution, the dashed ones for
            the more peaked A1 version, and the dotted ones for the less peaked
            A2 version.}
   \label{figvara}
\end{figure}

\begin{figure}
   \centering
   \includegraphics[width=9cm]{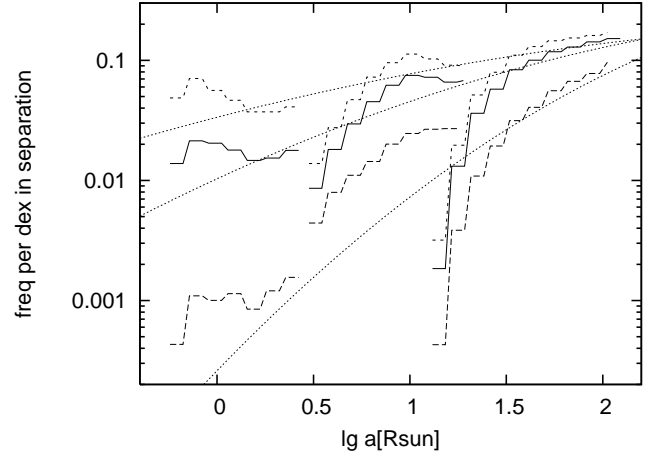}
   \caption{The number of close binary systems as calculated with the standard $f(a)$ (full lines), with
            the A1 modification (long dashes) or with the A2 modification (short dashes). The
            dotted lines show the input distributions, and relevant
            parts of the data for the ML(left),FG(middle) and O(right) bins are shown.
           }
   \label{figcmpa}
\end{figure}
\begin{figure}
   \centering
   \includegraphics[width=9cm]{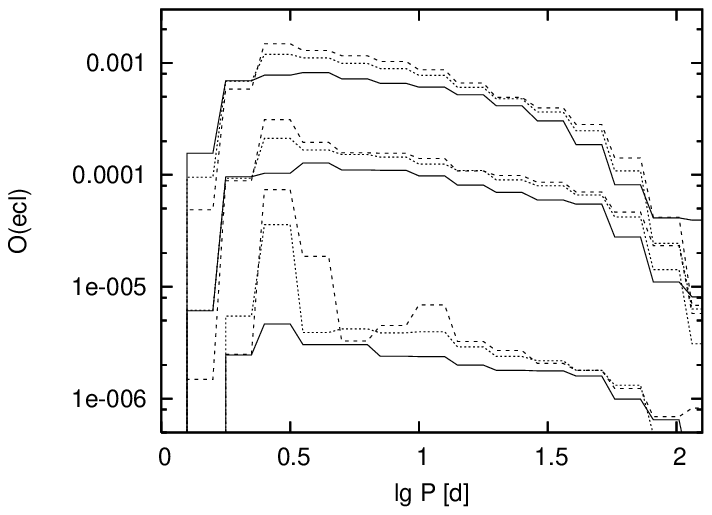}
   \caption{The $O_e$-curves for $\Delta m>0.1$, 0.4 and 0.8, for the O bin, for three different metallicities.
   The full curve is for a mean [Fe/H] about -0.55, the dashed line for +0.15, and the dotted line for
   [Fe/H]=-0.1.
           }
   \label{figcmpzo}
\end{figure}

\begin{figure}
   \centering
   \includegraphics[width=9cm]{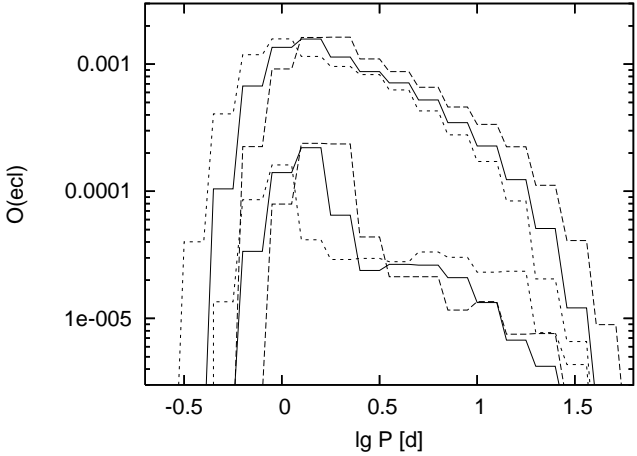}
   \caption{The $O_e$-curves for $\Delta m>0.1$ and 0.8, for the B bin, for three different metallicities.
   The full curve is for a mean [Fe/H] about -0.4, the dashed line for +0.2, and the dotted line for
   [Fe/H]=-1.0.
           }
   \label{figcmpz1}
\end{figure}

In order to see the differential effects of changes in $f(a)$ and
$f(q)$, it is sufficient to use only a single (solar) metallicity,
and also to use slightly incomplete samplings of the absolute
magnitude bins. With these simplifications, we have made 5 extra
sets of simulations at a number of absolute magnitude bins.
Instead of the nominal a-distribution in eq. (4), we have tried a
more peaked (A1), or a more uniform (A2) one. Instead of the
nominal q-distribution in eq. (3), we have tried either the
Duquennoy \& Mayor (\cite{Duquennoy}) (Q1) or a uniform (Q2)
$f(q)$, neither showing any peak at q=1. To demonstrate the
untenability of a random pairing q-distribution (cf. Malkov and
Zinnecker \cite{Malkov}), we also made a few runs (Q3) where the
secondary was chosen from the same IMF as the primary, producing
an $f(q)$ rising sharply towards small $q$. The assumptions are
summarized in Table \ref{runtab}. At each of the selected MS bins,
we have then four or five extra sets of $P_e/Q_e$-results apart
from the nominal one. Because we could not make as many
simulations as in the main runs, the numerical uncertainties are
larger, but the uppermost ($\Delta m>0.1$) curves are
well-defined.

\begin{table}
\begin{centering}
\begin{tabular}{|c|c|c|}
\hline
 run & $a$-distr & $q$-distr \\
\hline
 {\bf Std} & eq.(4)          &  eq.(3)           \\
{\bf A1} &   $n(\lg a-1.5,1.0)$    &  eq. (3) \\
{\bf A2} &   $n(\lg a-1.5,2.0)$    &  eq. (3)    \\
{\bf Q1} &   eq.(4)          & n(q-0.23,0.42) \\
{\bf Q2} &   eq.(4)          & uniform \\
{\bf Q3} &   eq.(4)          & random pairing \\
\hline
\end{tabular}
\caption{The differential runs, varying the a- or the
q-distributions. }\label{runtab}
\end{centering}
\end{table}

As expected, the $P_e$-data are usually less affected than the
$O_e$-data, because they depend mostly on the (similar) physical
characteristics of the individual components. The $O_e$-data
depend also on the close double period-distribution, which may
change appreciably with the input $f(a)$ and $f(q)$. As a simple
illustration, we may again look at the FG bin. Fig \ref{figvarq}
shows the 0.1 mag and 0.8 mag $O_e$-curves for the nominal run and
with the Q1-Q3 modifications. Small (Q1/Q2) changes in the
q-distribution are seen to have a rather small impact on the
$O_e$-curves, with the flat $f(q)$ giving marginally more, and the
DM one marginally fewer eclipsing binaries. As expected, however,
the Q3 model gives much fewer (about a factor of 5) eclipsing
binaries, even with an assumed 100\% duplicity. As the primary
mass increases, the predominance of small-mass secondaries gets
more and more extreme, producing very few similar-mass binaries
and less close binary interaction. For the OB-bin, a Q3 run gives
at least 50 times fewer eclipsing binaries than the standard ones,
clearly showing the problem faced by random pairing schemes in
creating sufficient numbers of eclipsing systems.

Fig \ref{figvara} shows the similar differential effects for the
A1 and A2 modifications. Here, the A1 curve is much lower, and the
A2 one clearly above the nominal. From similar calculations and
plots along the main sequence, this seems part of a regular trend,
with the largest effects at L and the smallest at O. Somewhat
surprisingly, these f(a)-effects are close to a simplistic
expectation looking only at the original f(a)-distributions. Fig
\ref{figcmpa} shows the raw number of systems for the three
different $f(a)$-assumptions, and for three different
mass-regions. To first order, the initial and final
period-distributions are proportional to each other, with the same
evolutionary details preserved. The final $O_e$-results do of
course include also the changes in $P_e$, actually turning the
somewhat small A1-effect for the FG stars in Fig \ref{figcmpa} to
the larger one in Fig \ref{figvara}.

As for the effects of metallicity, they vary in a complex way
along the main sequence. For many bins, we could compare two
partial runs with [Fe/H] $\sim$ +0.15 and [Fe/H] $\sim$ -0.55, and
generally, the $O_e$-values are higher for the higher metallicity.
In some simple cases, such as for the FG and K bins, the net
effect is mainly a shift of $\lg O_e$ in proportion to [Fe/H],
with a proportionality constant about 0.10 (per dex in
metallicity) at FG or 0.20 at K. For less massive stars, there is
also a change-of-shape, with higher $O_e$ for higher metallicity
at the long-period edge, but equal of fewer systems at short
periods. For the more massive stars, there is still a general
rise, but then also a more pronounced effect on the Algol-type
bump. This is both shifted towards longer periods, and increased
in amplitude with a rise in metallicity. For O-stars, the general
rise is of the order of 0.25, but the probability for 3-day
eclipsing binaries is very much dependent on [Fe/H], as can be
seen in Fig \ref{figcmpzo}. For the OB bin, the amplitude-effect
is smaller, but we have an obvious period-shift by about 0.2 in
$\lg P$ (per dex in [Fe/H]).

At B, the period-shift seems maximal, and in view of the
interesting observational results for the LMC treated in Sect. 8,
we made some extra runs with a larger range of metallicity. The
BSE-program does not allow z-values above 0.032, and the high-z
comparison is for a mean [Fe/H]=0.21. As a low-z sample, we used a
mean [Fe/H]=-1.0, and for an in-between set, a run with mean
[Fe/H]=-0.4 was already available. Fig \ref{figcmpz1} shows again
the $O_e$-curves for $\Delta m>0.1$ and $\Delta m>0.8$. There
seems to be a rather linear shift towards longer periods with
higher metallicity, by about 0.25 in lg P per dex in metallicity.
At A, the period-shift is down to about 0.1, and the
amplitude-shift to only some 0.04.

\section{Comparison with Hipparcos data}
There are few data sets where our theoretical curves can be
compared with observations. A crude but reassuring check can
however be obtained from the data in the Hipparcos Catalog (ESA
\cite{ESA1}). The basic assumption is that most of the eclipsing
binaries with periods in the 0.2 to 20 day range and
eclipse-depths larger than 0.1 magnitude were detected in the
Hipparcos photometry. Apart from the eclipse detection issues (see
below), the general completeness issues for Hipparcos are too
complex to be taken rigorously into account. For the present crude
check, we simply assumed a magnitude completeness limit $V=8.0$ at
galactic latitudes lower than 30 degrees, and $V=9.0$ at higher
latitudes. This bisection of the sky was chosen as a simplistic
first step towards a more realistic sampling, but in order to get
more than a few tens of eclipsing systems, we may have pushed the
completeness below 50\% at some latitudes (cf. Mignard
\cite{Mignard}). As long as the incompleteness is not correlated
with the fraction of eclipses, this should however not be a big
problem.

Using the Celestia software (ESA \cite{ESA2}), we then simply made
eclipse-statistics for two samples, one with parallaxes larger
than 4.5 mas and the other with parallaxes in the interval 1.0 to
4.5 mas. On average, the more distant sample refers to brighter
stars, and the aim of the division is to see some indication of
the trend with absolute magnitude. The observed (small!) numbers
were normalized by division by the total number of stars in the
same range of V and parallax (22585 for the distant sample and
34335 for the closer one), giving observational ratios to be
compared with our theoretical $O_e$-data.

\begin{figure}
   \centering
   \includegraphics[width=9cm,angle=0]{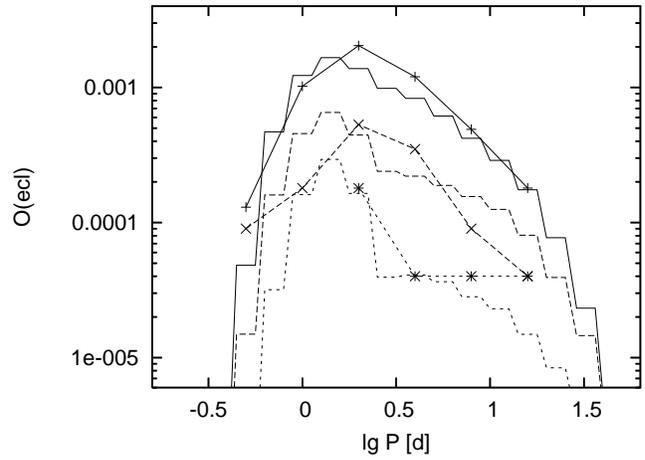}
      \caption{The observed numbers of binaries in the distant Hipparcos sample, together
      with the theoretical (B) $O_e$-curves. From above, the data are for $\Delta m >$ 0.1, 0.4 and
      0.7 magnitudes.}
         \label{fig_hipp1}
   \end{figure}

\begin{figure}
   \centering
   \includegraphics[width=9cm,angle=0]{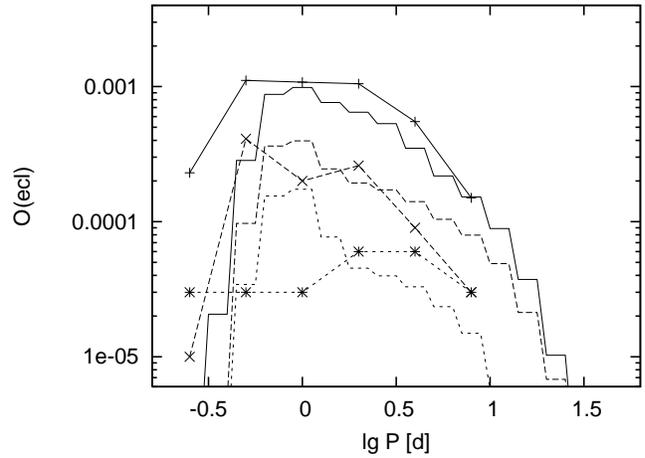}
      \caption{The observed numbers of binaries in the close Hipparcos sample, together
      with the theoretical (A) $O_e$-curves. From above, the data are for $\Delta m >$ 0.1, 0.4 and
      0.7 magnitudes.}
         \label{fig_hipp2}
   \end{figure}

To make the comparison, we have to know the distribution of
absolute magnitudes in the two samples. The $M_V$-distributions
are quite wide (10/90\% percentiles at -2.2/+0.9 for the distant
one, +0.3/4.0 for the closer one), but they differ enough to
influence appreciably the eclipse statistics. The median
magnitudes are -0.35 and 1.98, and comparing with Table 1, they
are not too far from the theoretical medians for the B and the A
bins. In view of the the many other uncertainties, no further
refinement seems warranted, and Fig \ref{fig_hipp1} and
\ref{fig_hipp2} show the Hipparcos data together with the
$O_e$-curves. The agreement is good for the more luminous sample,
while for the nearby sample, there seems to be more short-period
pairs observed than predicted. This is small-number statistics
(one system per bin at $\Delta m$=0.7), however, and the
differences are not significant. The general agreement of the
maximum $O_e$-levels is however an indication both that our input
f(a) is reasonable {\it and} that the binary fraction is high.
(Cf. Sect. 6).


Apart from a level-uncertainty due to the assumed binary fraction,
our theoretical curves also use the standard discovery criterion
with $\Delta m>0.1$ for 5\% of the orbit. This is admittedly
rather ad hoc, but it gives the reasonable large-period slope
shown in Fig. \ref{fig_hipp1}. Requiring only 0.05 mag eclipses
but for 10\% of the orbit pushes the long-period part of the
$O_e$-curves down as compared with the observations, and 0.1
mag/10\% even more so.

\section{Comparison with LMC data}
In recent years several surveys, e.g. MACHO, have been made
looking for microlensing events to test the hypothesis that a
significant fraction of the dark matter in the halo of the Milky
Way is made up of objects like brown dwarfs or planets. These
surveys are a gold-mine for all studies of variable stars, e.g.
eclipsing binaries. Several studies have been made from the data
produced by MACHO, and in Alcock et al. (\cite{Alcock}), the data
from the study towards the LMC has been analyzed. They extracted
there a sample of 611 eclipsing binaries by examining the light
curves of all potential short-period variables. We have analyzed
these data further regarding eclipse-depth and period, and we may
thus compare our theoretical studies with this observed sample.

Looking at the color-magnitude plot in Alcock et al.
(\cite{Alcock}), the observations seem reasonably complete for
apparent V magnitudes brighter than 18.0. Using a distance modulus
of 18.5 (Alves \cite{Alves}), plus a schematic 0.5 mag absorption
(cf. Zaritsky \cite{zaritsky}), this corresponds to absolute
magnitudes brighter than -1.0, that is somewhere between our OB
and B bins. Because the shape of the theoretical curves do not
vary fast with absolute magnitude, and in order to get a
reasonably large sample, we made P/$\Delta m$ statistics for all
systems with V$<$18, V-R$<0.3$. (This color-cutoff should
correspond roughly to our theoretical V-I$<$0.5, allowing for some
slight reddening).

For the theoretical comparison, we first interpolate our OB and B
results to an [Fe/H] to around -0.34 (Luck et al. \cite{Luck})
typical for young LMC-stars. We then use the mean absolute
magnitude in our observed sample (-2.2) plus the mean absolute
magnitudes given in Table 1 to interpolate again to mean
theoretical $O_e$-curves.
\begin{figure}
   \centering
   \includegraphics[width=9cm]{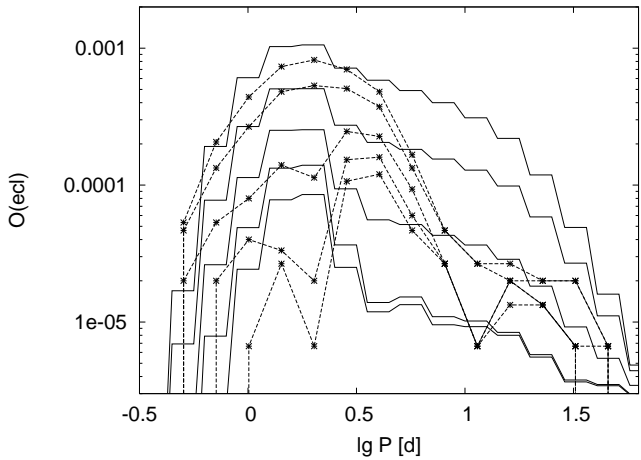}
   \caption{The distribution of eclipse depth as function of
            period for OB-stars in the LMC. The solid curves
            are our theoretical $O_e$ predictions for $\Delta m$=0.2, 0.4, 0.6, 0.8, and 1.0.
            The dashed lines are the observed MACHO data (arbitrarily shifted in the y direction).}
               \label{plotall_LMC}
\end{figure}

Since the LMC data are not normalized, the observed histograms can
be shifted arbitrarily in the y-direction, but from Fig
\ref{plotall_LMC}, we see at once that the observations put the
maximum in the period-distribution for deep eclipses at a much
longer period than expected from the theory. There is the
predicted hump, but at the wrong period. A similar shift was found
in Sect. 6, but the magnitude is much too high to be explained
even by super-metal-rich stars. The large-scale variations in
$f(a)$ and $f(q)$ as explored in Sect. 6 also do not give a shift
of the period-maximum, and we have to make sure that we do not
overlook any possible systematic effects in the observations.

A first necessary test is of course to try to see any changes in
the observed curves with respect to absolute(=apparent) magnitude,
noting the theoretical prediction of longer hump periods for
brighter stars. No such effects could be seen for different
magnitude and color cutoffs (and only 5 out of 563 systems are
bright enough to be in our O bin). It is also clear that many of
these close binaries show large out-of-eclipse light-variations.
Because we use the full max-min amplitude, we overestimate the
eclipse-depths, and more so the shorter the period. This will in
effect aggravate the fit to the theory, depressing the observed
curves at short periods which are already too low. Similarly, the
detection completeness should be best at short periods, again
going against the observed effect. The only slight correction from
longer to shorter periods would be where a short-period system
with small secondary minimum is miss-classified as a double-period
system with equal minima. Such cases may occur, but not in enough
numbers to make a qualitative change.

Although it seems very plausible that some of the internal BSE
parameters governing close binary evolution need adjustment, it is
hard to see why these adjustments should be so different between
LMC and the Galaxy. The reason for the bad fit in Fig
\ref{plotall_LMC} is thus unclear, but some small-scale
peculiarities in the birth f(a)-distribution and/or the
interstellar environment (influencing the mass-loss/mass-transfer
processes) may perhaps be involved.

\section{Conclusions and directions for future studies}
There are a number of conclusions to be drawn from the present
investigation. The most obvious one is of course that for
predictions about eclipsing binaries, one has to allow for the
full complexity of close binary evolution, and that only with the
availability of a rapid interpolation tool like BSE, are
multi-million system population syntheses at all feasible. In this
first study, we have used BSE with standard parameters, but for
increased realism, some of them should be optimized further (see
below).

Our main result is a theoretical prediction about the periods and
depths of main sequence eclipsing binaries, as a function of their
absolute magnitude. An admittedly crude comparison with real
Hipparcos observations shows a quite reasonable fit for the number
of eclipsing binaries, and even for their distribution over period
and eclipse depth. Encouraged by this fit, we have tried also to
estimate the properties of the eclipsing binaries observable by
ESA's Gaia mission, see Dischler \& S\"oderhjelm (\cite{diso05})
and S\"oderhjelm (\cite{sod05}). For B-stars in the LMC, we could
only look at the $P/\Delta m$ distributions, which is however
sufficiently offset from the theoretical one that one is led to
assume some rather different input orbit-distributions than in the
Galaxy.

As a general result, we have concluded that the number of observed
eclipsing binaries scales rather directly with the input frequency
of short-period orbits, although the post-evolution
period-distribution is far from the input one. The good fit for
the Hipparcos samples shows that at least in the short-period end,
the simple log-normal Duquennoy-Mayor a-distribution (as derived
originally for G-stars) is a good model even for the upper main
sequence. As for the mass-ratios, the present results (not
unexpectedly) rule out all q-distributions with an excess of
small-mass secondaries, as obtained e.g. by choosing the primary
and secondary masses from the same IMF. For flatter distributions,
small details (like the $q=1$ peak) turn out to have rather little
influence on the observable eclipsing-binary population.

Already from our existing data, we may go on to derive theoretical
distributions of spectroscopic and visual binaries in the same
main-sequence result-bins. Comparisons with observed data will
then allow conclusions about $f(a)$ over a much larger a-interval.
To improve the results for the eclipsing binaries, the obvious
next step is to calculate more realistic light-curves, with
outside-eclipse variations and limb-darkening. Also, with enough
computer resources, it should be interesting to perform
large-scale studies where also some of the internal BSE parameters
are varied, and the results should be given for smaller areas of
the HR-diagram.

\section*{Acknowledgement}
We thank Jarrod Hurley for giving us free access to the crucial
BSE-code, Chris Tout for some advice as to how to use it, and Onno
Pols for a thorough refereeing of the final article.

{}

\clearpage

\appendix

\section{Sample eclipse probabilities} Below are given two sample $P_e/Q_e$
tables, giving the data for $\Delta m$=0.1 and 0.8 at 18 different
periods (spacing 0.15 in lg P). Each row corresponds to a MS bin
(with brown dwarfs in the ML+L combination). The column headings
give the mean period in days, and in the A.1 table are given the
corresponding mean $\lg P_e$ values. In the A.2 table, we give
instead $\lg O_e$, where $O_e$ is the fraction of eclipsing
binaries (per 0.3 dex in period) relative to all stars in this
bin, assuming 80\% binarity in our population synthesis model.


\subsection{$P_e$-values}

\noindent
{\scriptsize
\begin{tabular}{|c|cccccccccccccccccc|}
\hline
& \multicolumn{18}{|l|}{$\Delta M \geq 0.1 $} \\
\hline
Bin &0.175 &0.25 &0.36 &0.51 &0.72 &1.02 &1.44 &2.04 &2.88 &4.08 &5.76 &8.16 &11.5 &16.3 &23.0 &32.5 &46 &65    \\
\hline
O&-&-&-&-&-&-&-0.50 & -0.39 & -0.55 & -0.79 & -1.00 & -1.13 & -1.25 & -1.39 & -1.54 & -1.71 & -1.91 & -2.30 \\
OB &-&-&-&-0.22 & -0.31 & -0.36 & -0.44 & -0.55 & -0.74 & -0.92 & -1.04 & -1.21 & -1.41 & -1.61 & -1.89 & -2.29 & -2.74 & -3.31 \\
B & -0.87 & -1.00 & -0.74 & -0.43 & -0.42 & -0.49 & -0.63 & -0.83 & -1.05 & -1.16 & -1.32 & -1.53 & -1.74 & -2.00 & -2.39 & -2.94 & -3.76 & -4.30 \\
A & -0.92 & -0.95 & -0.79 & -0.68 & -0.69 & -0.84 & -1.04 & -1.17 & -1.26 & -1.46 & -1.69 & -1.89 & -2.16 & -2.59 & -3.18 & -4.13 & -5.61 &  - \\
FG& -1.05 & -0.89 & -0.87 & -0.95 & -1.05 & -1.18 & -1.29 & -1.49 & -1.75 & -1.98 & -2.25 & -2.70 & -3.32 & -4.30 & -5.92 & - & - & - \\
K& -0.93 & -0.91 & -0.95 & -1.03 & -1.31 & -1.63 & -1.77 & -1.92 & -2.17 & -2.59 & -3.57 & - & - & - & - & - & - & - \\
M(a)& -0.70 & -0.86 & -1.15 & -1.37 & -1.44 & -1.60 & -1.81 & -2.16 & -2.92 & -4.48 & - & - & - & - & - & - & - & - \\
M(b)& -0.78 & -1.16 & -1.61 & -1.73 & -1.86 & -2.16 & -2.58 & -3.64 &  -    &   -   & - & - & - & - & - & - & - & - \\
ML+L & -1.00  & -1.45 & -1.80 & -2.15 & -2.56 & -3.31 & -4.35 & -6.41 & - &   -   & - & - & - & - & - & - & - & - \\
\hline
& \multicolumn{18}{l|}{$\Delta M \geq 0.8 $} \\
\hline
O  & - & - & - & - & - & - & - & -3.32 & -2.28 & -3.64 & - & - & -4.21 & - & - & -5.55 & -5.49 & -5.15 \\
OB & - & - & - & - & - & -2.42 & -1.77 & -1.59 & -1.97 & -3.26 & -3.09 & -3.28 & -3.26 & -3.23 & -3.46 & -3.43 & -3.43 & -3.62 \\
B & -1.35 & -2.21 & -1.80 & -2.55 & -1.81 & -1.51 & -1.47 & -1.84 & -2.58 & -2.57 & -2.65 & -2.81 & -2.95 & -3.17 & -3.42 & -3.81 & -4.46 & -4.39 \\
A & -1.42 & -1.52 & -1.96 & -1.81 & -1.58 & -1.70 & -2.19 & -2.53 & -2.59 & -2.70 & -2.91 & -3.23 & -3.94 & -5.52 & -6.33 & -6.58 & -6.76 & - \\
FG & -1.23 & -1.45 & -1.87 & -2.10 & -2.39 & -2.83 & -3.01 & -3.22 & -3.45 & -3.69 & -4.45 & - & - & - & - & - & - & - \\
K & -1.22 & -1.53 & -2.11 & -2.26 & -2.76 & -3.06 & -3.18 & -3.14 & -3.21 & -3.51 & -4.45 & - & - & - & - & - & - & - \\
M(a) & -2.65 & -3.09 & -3.68 & -4.44 & -4.76 & -5.04 & -5.10 & -5.57 & - & - & - & - & - & - & - & - & - & - \\
M(b) & -2.36 & -2.59 & -2.96 & -3.24 & -3.40 & -3.71 & -4.40 & -5.21 & - & - & - & - & - & - & - & - & - & - \\
ML+L & -1.38 & -1.77 & -2.15 & -2.57 & -3.08 & -3.92 & -4.86 & -7.39 & - & - & - & - & - & - & - & - & - & - \\
\hline
\end{tabular}
}


\subsection{$O_e$-values}

\noindent
{\scriptsize
\begin{tabular}{|c|cccccccccccccccccc|}
\hline
& \multicolumn{18}{|l|}{$\Delta M \geq 0.1 $} \\
\hline
Bin &0.175 &0.25 &0.36 &0.51 &0.72 &1.02 &1.44 &2.04 &2.88 &4.08 &5.76 &8.16 &11.5 &16.3 &23.0 &32.5 &46 &65    \\
\hline
O & - & - & - & - & - & - & -4.00 & -3.17 & -2.93 & -2.96 & -3.01 & -3.06 & -3.12 & -3.22 & -3.33 & -3.45 & -3.62 & -3.97 \\
OB & -& -&  -& -5.95& -4.16& -3.29& -2.91& -2.77& -2.87& -3.00& -3.08& -3.21& -3.36& -3.50& -3.72& -4.05& -4.49& -5.04 \\
B & -6.59& -6.87& -5.97& -4.32& -3.33& -2.91& -2.78& -2.86& -3.01& -3.08& -3.21& -3.38& -3.54& -3.76& -4.11& -4.63& -5.42& -5.94 \\
A & -5.82& -5.87& -4.69& -3.55& -3.06& -3.01& -3.12& -3.19& -3.27& -3.46& -3.66& -3.82& -4.05& -4.43& -4.99& -5.92& -7.37& - \\
FG & -5.33& -4.78& -4.24& -3.87& -3.67& -3.57& -3.51& -3.57& -3.74& -3.98& -4.30& -4.75& -5.32& -6.26& -7.84& -& -& - \\
K & -5.18& -4.83& -4.51& -4.20& -4.13& -4.11& -4.02& -4.05& -4.31& -4.74& -5.70& -& -& -& -& -& -& - \\
M(a) & -3.94& -3.86& -3.93& -3.98& -3.93& -4.08& -4.24& -4.52& -5.22& -6.71& -& -& -& -& -& -& -& - \\
M(b) & -3.65& -3.80& -4.09& -4.18& -4.38& -4.70& -5.05& -6.07& -& -& -& -& -& -& -& -& -& - \\
ML+L & -4.09& -4.20& -4.37& -4.67& -5.17& -5.94& -6.92& -8.91& -& -& -& -& -& -& -& -& -& - \\
\hline
& \multicolumn{18}{l|}{$\Delta M \geq 0.8 $} \\
\hline
O & -& -& -& -& -& -& -& -6.10& -4.66& -5.81& -& -& -6.08& -& -& -7.29& -7.20& -6.81 \\
OB & -& -& -& -& -& -5.35& -4.23& -3.81& -4.10& -5.35& -5.13& -5.27& -5.21& -5.12& -5.29& -5.19& -5.18& -5.34 \\
B & -7.08& -8.07& -7.03& -6.44& -4.72& -3.93& -3.62& -3.86& -4.55& -4.49& -4.54& -4.65& -4.75& -4.93& -5.15& -5.50& -6.12& -6.04 \\
A & -6.32& -6.44& -5.86& -4.67& -3.95& -3.87& -4.26& -4.55& -4.61& -4.70& -4.87& -5.15& -5.83& -7.36& -8.14& -8.36& -8.52& - \\
FG & -5.52& -5.34& -5.24& -5.02& -5.00& -5.22& -5.23& -5.30& -5.44& -5.70& -6.50& -& -& -& -& -& -& - \\
K & -5.47& -5.46& -5.66& -5.43& -5.58& -5.55& -5.43& -5.27& -5.36& -5.66& -6.58& -& -& -& -& -& -& - \\
M(a) & -5.89& -6.09& -6.47& -7.04& -7.24& -7.52& -7.53& -7.93& -& -& -& -& -& -& -& -& -& - \\
M(b) & -5.24& -5.24& -5.43& -5.69& -5.92& -6.25& -6.87& -7.64& -& -& -& -& -& -& -& -& -& - \\
ML+L & -4.47& -4.53& -4.73& -5.09& -5.69& -6.55& -7.43& -9.89& -& -& -& -& -& -& -& -& -& - \\
\hline
\end{tabular}
}

\end{document}